\documentstyle[aps,prl,epsf,multicol]{revtex}
\begin{document}
\draft
\title{ Transport Properties of the One Dimensional Ferromagnetic
Kondo Lattice Model : A Qualitative Approach to Oxide Manganites}
\author{J.J. Betouras$^1$ and S. Fujimoto$^{1,2}$}
\address{
$^{1}$Department of Physics, Theoretical Physics,
University of Oxford,
1 Keble Road,
Oxford, OX1 3NP,
U.K.\\
$^{2}$Department of Physics, Kyoto University, Kyoto 606, Japan}
\date{today}
\maketitle

\begin{abstract}
The transport properties of the ferromagnetic Kondo lattice model in one
 dimension are studied via bosonization methods. The antiferromagnetic
fluctuations, which normally appear because of the RKKY interactions, 
are explicitly taken into account as a direct exchange between the
 ``core'' spins. It is shown that in the paramagnetic regime 
with the local antiferromagnetic fluctuations,
the resistivity decays exponentially as the temperature increases
while in the ferromagnetic regime the system is an almost perfect
conductor.
 The effect of a weak applied field is discussed to be reduced to the case
of the 
ferromagnetic state leading to band splitting. 
The qualitative relevance of the results for the problem of the Oxide
Manganites is emphasized. 
\end{abstract}

\begin{multicols}{2}
The Kondo lattice model fundamentally captures the physics of heavy
electron systems (e.g. Uranium or Cerium compounds). Therefore the
efforts to study it have been intensive for a long time. One route
to the full understanding has been the study of the one-dimensional(1D)
analogue due
to well developed techniques that can be used \cite{tsunetsuguueda}.
More recently another class of materials, where a ferromagnetic version of the
Kondo lattice model is expected to be the starting point, has been 
extensively studied.
These are the manganite oxides of the form Ln$_{1-x}$A$_{x}$MnO$_{3}$ 
where Ln is a lanthanide and A is an alkaline-earth element, which
possess a very rich phase diagram \cite{experiment1}. As a function of
doping there are regions of ferromagnetic (FM), antiferromagnetic (AFM)
as well as charge ordering. Moreover there is an almost insulating state
above the FM critical temperature, where a huge increase in
resistivity is observed and becomes even more dramatic with the
application of an external field. This phenomenon is 
termed as ``Colossal Magnetoresistance'' (CMR).
The FM order can be explained by the double exchange mechanism \cite{anderson}.
This mechanism though, is clearly not enough to explain the CMR 
\cite{millisprl} and other possibilities have
been proposed - e.g. the relevance of Jahn-Teller
distortion \cite{millis2}, the effect of AFM
fluctuations or localization due to randomness etc. \cite{moreo} 
- to explain the phenomenon. A deep 
understanding of the  relevant electronic models is vital and the FM Kondo
Model, especially in 1D case, 
is the first candidate in this direction as well.

Recently Monte Carlo calculations have demonstrated the potential of a
very rich phase diagram of the FM Kondo Hamiltonian
\cite{dagotto1}. At low temperatures and as a function of
doping and interaction strength, regimes with paramagnetic, FM,
incommensurate correlations and also phase separation
between undoped AFM and hole-rich FM regions have been observed. The
similarities in the behavior of the 1D system with a higher
dimensional one, are also emphasized.

In this work a reversed question is examined: given the local spin
configuration, what the consequences for the transport properties are.
The starting point is the following Hamiltonian  :
\begin{equation}
H= - t \sum_{<i,j>,\alpha} {c}_{i,\alpha}^{\dagger}
c_{j,\alpha} + h.c.- J_H
\sum_{i,\alpha,\beta} {c}_{i,\alpha}^{\dagger}\sigma_{\alpha,\beta}
c_{i,\beta} \cdot \vec{S_i}.
\end{equation}
In the ``language'' of the CMR problem, the first term represents
the $e_g$ electron hopping between nearest-neighbor Mn ions at sites
i and j. The second term is the Hund's rule coupling between the 
localized spins $\vec{S_j}$ (the $t_{2g}$ singly occupied orbitals
with S=3/2) and the mobile $e_g$ electron at the same site.
A direct AFM exchange between the nearest  ``core'' spins
$H_{AFM}= J_{AFM} \displaystyle\sum_{<i,j>} \vec{S_i} \cdot \vec{S_j}$
can be added to ensure the presence of the AFM fluctuations. This
effect is present and important at half filling,
even in the absence of the direct exchange term, through the RKKY
interactions. The latter are the effective interactions between the ``core''
spins once the fermionic degrees of freedom are integrated out.
However the need of this direct exchange term has to be emphasized on
pure experimental grounds since AFM short range order has been
observed by neutron scattering experiments on
${\rm La_{1.2}Sr_{1.8}Mn_2O_7}$ (which is a compound of reduced
dimensionality) \cite{perring} in addition to the fact that regimes
of AFM ordering are adjacent to the FM as a function of doping in
the experimental phase diagram.
Moreover the direct exchange interaction is always present \cite{rama} and
can be crucial for the formation of localized spin polarons,
and for the insulating behavior in the paramagnetic phase.
Thus, we take into account this direct AFM interaction
explicitly,
and investigate the effects of local AFM fluctuations.
Following the experimental observation\cite{perring}, 
we assume $J_{AFM}\ll J_H$.
The opposite limit $J_{AFM} \gg J_H$, the case of which
was studied before in another context,
gives a quite different physics\cite{dop}.

The approach followed in this Letter is the abelian and non-abelian
bosonization \cite{affleck} of the charge and spin
degrees of freedom of the conduction electrons respectively, 
with the coherent path integral method for the localized spins. 
Then
the effective action reads :
\begin{eqnarray}
\nonumber
S&=& \int dxd\tau [v (\partial_x \phi_c)^2 + \frac{1}{v} (\partial_{\tau}
\phi_c)^2] + S_{WZW} (g) \nonumber \\ 
&&+J_{Hr}\int dx d\tau (\vec{J_L} + \vec{J_R} + e^{i2k_{F}x}
{\rm Tr}(g\vec{\sigma}) e^{i\sqrt{2\pi} \phi_c} \nonumber \\
&&+ h.c. ) \cdot \vec{S_i} +S_{AFM},
\end{eqnarray}
where $\phi_c$ is the boson field for the charge degrees of freedom,
$\vec{J}_L$ and $\vec{J}_R$ are the spin current operators which satisfy level-1
SU(2) Kac-Moody algebra, $v$ is the velocity which is the same for the charge and
spin degrees of freedom, g(x) is a representation of
SU(2) and $S_{WZW}$ is the Wess-Zumino-Witten action given by \cite{tsvelik}:
\begin{eqnarray}
S_{WZW}&=& -\frac{1}{16 \pi} \int d^2 x
{\rm Tr}(\partial_{\mu}g^{-1}\partial_{\mu}g) \nonumber \\
+\frac{i}{24 \pi}&& \int_{0}^{\infty}d \xi \int d^2 x
\epsilon^{\alpha\beta\gamma} {\rm Tr}(g^{-1}\partial_{\alpha}g
g^{-1}\partial_{\beta}g g^{-1}\partial_{\gamma}g).
\end{eqnarray}  
$S_{\rm AFM}$ is the action for the localized spins
obtained by applying the coherent path
integral method to $H_{AFM}$\cite{frad}.

In the absence of a direct exchange term, at half-filling, the $J_{Hr}$
term (which comes from the renormalization of $J_H$) is marginally
irrelevant and scales to zero. On the other hand the RKKY interaction
becomes important and gives rise to a dispersion of localized spins 
\cite{tsvelikprl,sfuji}. 
The configuration of the localized spins can be 
parametrized in the case of general filling as :
\begin{equation}
\vec{S_j}= S(\vec{l_j} + \sqrt{1-{l_j}^2} \vec{m_j}),
\label{sj}
\end{equation}
with j enumerating the cites and $\vec{l} \cdot \vec{m}=0$ where
$\vec{l}$ is the fast variable whereas a unit vector 
$\vec{m}$ is the slow one. 

{\bf{Paramagnetic regime}}: 
We first consider the effects of local AFM fluctuations
in the paramagnetic (PM) state.
We assume $J_{AFM}\gg J_H^2/v$.
This condition ensures that the direct AFM correlation
between localized spins dominates over
the $2k_F$ RKKY correlation.
This situation is actually realized in the systems
with the 2D layered structure, ${\rm La_{1.2}Sr_{1.8}Mn_2O_7}$,
for which a commensurate AFM correlation
is observed in the paramagnetic insulating state\cite{perring}.
If the electron density is relatively close to
half-filling, the tendency to the FM state is suppressed.  
This assumption is also consistent with 
the recent numerical results\cite{moreo,dagotto1}.
In this case, it is plausible to assume the presence of 
the finite AFM correlation length $\xi_{\rm AFM}$.
Then, within the scale smaller than $\xi_{\rm AFM}$,
we can put $\vec{m} = e^{-i\pi j} \vec{n}$ with $\vec{n}$ a unit vector.
By substituting the above expression into the action, we get from
the $J_H$ term the following expression :
\begin{eqnarray}
\nonumber
&&(\vec{J_L}+\vec{J_R}) \cdot \vec{l_j} + ({\rm Tr}(g\vec{\sigma}) e^{i
\sqrt{2\pi} \phi_c} e^{(2 k_F - \pi)x} + h.c.) \sqrt{1- l^2} \vec{n} \\ 
\nonumber
&&+(\vec{J_L}+\vec{J_R}) \sqrt{1- l^2} e^{-i\pi x} \vec{n} \\
&&+( {\rm Tr}(g\vec{\sigma}) e^{i2k_F x} 
e^{i \sqrt{2\pi} \phi_c} + h.c.) \vec{l}.
\end{eqnarray}
The first term proportional to $\vec{l}$ 
which carries the fluctuations of the spin
configuration
coupled with the degrees of freedom of the conduction electron is
not important for the transport properties which are of primary concern,
since it does not couple to the charge degrees of freedom. 
The last two terms which contain the rapidly oscillating factor
give subdominant contributions.
The second, proportional to $\vec{n}$ term is predominant and 
is going to generate a gap.
In order to deal with this term, we follow the method developed by
Tsvelik\cite{tsvelik}.
We first neglect the fast variable and approximate $\sqrt{1-l^2} \simeq 1$.
The effect of the fast variable will be considered later. 
By writing $i\vec{\sigma} \cdot \vec{n} \equiv h  \in$ SU(2) and
subsequently $gh \equiv G$ this term is transformed to :
\begin{equation}
({\rm Tr}(G) e^{i\sqrt{2\pi}\phi_c} e^{i(2k_F-\pi)x} + h.c.).
\end{equation}
By refermionizing the above expression we get ${\rm Tr}(G)
e^{i\sqrt{2\pi}\phi_c}= {\psi_{R,\alpha}}^{\dagger} \psi_{L,\alpha}$
where summation over spin indices is assumed. 
Then the part of the action which involves only the degrees of
freedom of  the conduction electron, together with the above term can
be refermionized into the form :
\begin{eqnarray}
\nonumber
S_0=\int dxd\tau [v{\psi_{R,\alpha}}^{\dagger}\partial \psi_{R,\alpha} -
v{\psi_{L,\alpha}}^{\dagger} \partial \psi_{L,\alpha} \\
+ J_HS
{\psi_{R,\alpha}}^{\dagger}\psi_{L,\alpha} e^{i\delta x} + h.c.]
\label{mix}
\end{eqnarray}
We have defined  $\delta=2k_F-\pi$.
The diagonalization of the above Hamiltonian is an easy task giving
the dispersion relation : $\epsilon_k = \pm \sqrt{v^2k^2+{J_H}^2{S}^2}
-v\delta/2$. The important feature of the above analysis is that when
$J_HS \geq v\delta/2$  a massive state with a gap $\Delta=J_HS-v|\delta|/2$
appears despite the fact
that an oscillatory factor exists in the crossed term of
Eq.(\ref{mix}). 
In this massive state, we should interpret $v|\delta|/2$ as a chemical
potential which lies within the gap.
The strong correlation between the gap and the distance
from half-filling is remarkable. In principle if $J_HS > v|\delta|/2$
there is always a gap in the spectrum.
Note that our approach is non-perturbative in $J_HS/v$.
This argument is consistent 
under the condition that the local AFM correlation length is large enough to
produce the mass gap; {\it i.e.} $\Delta^{-1}<\xi_{\rm AFM}$.
Following ref.\cite{tsvelikprl}, we can show that
the spin degrees of freedom are described
in terms of an $O(3)$ nonlinear $\sigma$ model without the topological term.
If the relation $J_{AFM}\gg J_H^2/v$, which is the condition
for the presence of local AFM fluctuations,
 holds, 
the mass gap of the spin excitation is given by 
$\sim \exp(-\pi S)$. 
Thus for sufficiently large $S$, the above condition,
$\Delta^{-1}<\xi_{\rm AFM}$,
is satisfied, and our argument is applicable. Note that the gap
$\Delta$ is not a function of $J_{AFM}$ but only of $J_H$ and
$\delta$. The direct AFM term is physically needed to ensure the existence of
the PM state in the form of short AFM correlations 
(even away from half - filling).

The dc conductivity for this state can
be easily computed using the Kubo formula (current-current correlation)
and
we have the expression for the resistivity:
$\rho \propto e^{\Delta/kT}$ for $kT\ll \Delta$. 
The physics behind these
calculations is that in the paramagnetic state the strong AFM fluctuations
are able to form bound states between the localized spins and the
conduction electrons, leading to a gap generation \cite{tsvelikprl}. 
The only possibility to achieve non-zero conductivity is through
temperature fluctuations which can ``break'' these bound states.
The insulating state is quite due to the frustration effect
of the spin systems.
The itinerant electrons give rise to
the frustration between $2k_F$ RKKY interaction and the direct 
exchange interaction.
In order to suppress this frustration, the localized state
of electrons is favored energetically for sufficiently large $J_H$.
 
So far we have neglected the fast variable $\vec{l}$.
Even if we include it, the above result is not changed essentially.
After the gap formation, taking up to the second order in $\vec{l}$,
we integrate out the fast variable.
Then, we find that it just renormalizes the mass gap, and
does not change the low-energy qualitative properties.


{\bf{Ferromagnetic regime}} : Now let's consider the FM state.
When $\delta$ increases,
the gap $\Delta$ obtained in the vicinity of the half-filling collapses.
This implies the backward scattering term
of $J_H$ term becomes irrelevant, and that
the forward scattering term of $J_H$, which induces the FM fluctuation
between localized spins, 
determines the low-energy physics,
overcoming the AFM fluctuations due to the direct exchange interaction.
This picture is again supported by the numerical results\cite{moreo,dagotto1}.
Although the FM state is realized for $J_H/t\gg 1$, in which case
the bosonization approach is not applicable,
we can stabilize the FM state for a small value of $J_H/t$
by applying an external magnetic field as will be discussed later.
Thus, here, we assume the presence of the FM order.
In this case, 
we can parametrize $\vec{S}=S\vec{n}=S(\sin \varphi\cos\alpha,
\cos\varphi\cos\alpha,\sin\alpha)$ in eq.(\ref{sj}),
where the averaged magnetization in the FM state 
is assumed to be in the $z$-direction. 
The action which describes the spin wave is given by 
\begin{equation}
S_{SW}=\int dx d\tau v_s[iS\alpha c^{-1}\partial_{\tau}\varphi
+(\partial_x\varphi)^2
+(\partial_x\alpha)^2], \label{sw}
\end{equation}
where $c$ is a constant.
The coupling between conduction electrons and the $\vec{n}$ 
field through $J_H$ term is given by,
\begin{equation}
J_H(\vec{J_L} + \vec{J_R})\cdot\vec{n} + J_HS(e^{i2k_{F}x}
{\rm Tr}(g\vec{\sigma}) e^{i\sqrt{2\pi} \phi_c} + h.c. ) \cdot \vec{n}. 
\label{fjh}
\end{equation}
The second term of the above equation 
which involves the charge degrees of freedom
contains a rapidly oscillating factor $\exp(2ik_Fx)$.
Thus, it is an irrelevant interaction at the low-energy fixed point,
and the system is described in terms of
free fermions in a magnetic field (Zeeman interaction) 
with the dispersion relation: 
$\epsilon_k = v\vert k\vert \pm J_HS$.
There is no finite resistivity at this low-energy fixed point.
The system is an ideal metal.
Since in 1D systems the transition temperature to the FM state
$T_c$ is equal to zero, we can not discuss about the resistivity
at finite temperatures in this framework.
However, it is an instructive and interesting issue to consider the temperature
dependence of the resistivity in this state assuming
$T_c\neq 0$ due to some magnetic coupling 
with three-dimensional directions \cite{john}.
In this case, the second term of eq.(\ref{fjh})
may give rise to the resistivity at finite temperatures.
In order to see this, 
we calculate the resistivity using the memory function formalism, and
treating the above term perturbatively as done by Giamarchi
for the Sine-Gordon model\cite{gia}.
 This approach for the calculation of
resistivity is followed with special care, because it gives a finite value
even for 1D integrable systems, which are
intrinsically ideal metals or ideal insulators\cite{zotos}. 
However, since our system is not integrable, 
we expect that this method gives qualitatively correct temperature
dependence of the resistivity.
The lowest order non-vanishing contribution comes from
the second order term in $J_HS$ after averaging over $\varphi$ 
and $\alpha$.
The oscillating factors $\exp(\pm4i(k_F/v)x)$ which appear 
in all terms involving the charge degrees of freedom
determine the dominant temperature dependence of the resistivity.
The result is given by $\rho\propto J_H^2S^2\exp(-a(k_Fv)/T)$
with $a$ a constant for $T \ll k_{F}v_{F}$.
Although there is a prefactor which depends on the temperature,
the dominant temperature dependence is determined by
the exponential decay. 
Thus, the resistivity decreases very rapidly in the FM state as
temperature decreases, and at zero temperature
the system is an ideal metal with an infinite conductivity.
This feature should be closely relevant to
the colossal change of the resistance observed in oxide manganites.

{\bf {Effect of an applied field}}: We now 
discuss the effect of an external field $H_0$ ($> H_c$) which is a direct
extension of the FM case. The qualitative
argument for this statement can be understood in two basic steps.
In the first place the application of a field is going to
reduce the effect of the antiferromagnetic fluctuations
and (through a Zeeman term)
to produce a non-zero average - both in time and space -
magnetization of the localized spins towards the preferred direction.
The second step is that through the strong $J_H$ term the ``effective
field'' which affects the conduction electron is enhanced by
the factor $J_H$ and is much stronger.
Therefore our calculations (definitely in the FM regime for 
$\mu H_0 > J_{AFM}S^2$, $\mu$ the magnetization  of the localized
spins) indicate 
again that the
two bands split and the energy difference is exactly what has been found in the
FM case enhanced by 2 $\vec{\mu} \cdot {\vec{H}}_{eff} = 2 \mu H_{eff}$,
where here $\mu$ is the magnetization of the conduction electrons. 
The application of an external field therefore largely contributes 
to the increase of the conductivity as observed in all the
experimental data. A more detailed study of the crossover and the
existence the critical field $H_c$ above which the FM behavior is
obtained, is a different issue that will be reported elsewhere. The
basic question that has to be addressed, in order to give a full
answer of this specific problem, is the precise effect of the
field in the renormalization of the mass gap in the paramagnetic state
or equivalently to take into account the intermediate states from the
PM to FM.

In summary, we have analyzed the transport properties 
of the 1D FM Kondo Model with AFM
fluctuations in two different regimes of the
phase diagram, discovering an almost insulating behavior in the
paramagnetic case and an almost zero resistivity in the FM case. 
As a general conclusion of these non-perturbative calculations we
would like to stress the interplay - by purely magnetic interactions - between
the ordered and the paramagnetic state. The very fact that numerical
calculations in the model are able to demonstrate the existence
of similar phases both in 1d 
and in higher dimensions, in agreement with the
available experimental data, serves as a strong indication that the
model and the results are relevant to the CMR problem. Since in 1D 
there is not an actual phase transition at finite temperatures, 
we lack a detailed description
relevant to the temperature range in the vicinity of the actual
$T_c$ (besides the totally different behavior in the two regimes that has been
discussed). The physical picture we would like to emphasize 
though, is that given a fixed set of $appropriate$ interaction
parameters 
and filling, the system
can be in the FM or in the PM regime as a function of
temperature,
with the consequences described
above which are qualitatively very close to the experimental findings.
Therefore, the basic idea 
of  this mechanism can be extended to more than 1D and,
in addition to the strong electron-lattice coupling or other 
mechanisms (which are of
secondary importance in this approach), may explain the dramatic
dependence of the resistivity on the temperature.

We are grateful to Elbio Dagotto, Fabian Essler and especially to
John Chalker and Alexei Tsvelik
for many discussions and suggestions on this
work. We also wish to thank Stephen Blundell for 
discussions on the
experimental aspect of the CMR. J.B. would like to thank 
Robert Joynt, Andrey Chubukov, Dirk
Morr and Mark Rzchowski for relevant discussions and
to acknowledge the financial support of the European Union through the
Marie Curie Fellowship.
S.F. would like to thank the Department of Physics,
University of Oxford for the hospitality and acknowledges financial support
from the Ministry of Education, Science, and Culture of Japan.

\end{multicols}
\end{document}